# Hide the Image in FC-DenseNets to another Image

Duan xintao[1], Liu Nao[1]. [1]HeNan normal University

**Abstract**：In the past, steganography was to embed text in a carrier, the sender Alice and the recipient Bob share the key, and the text is extracted by Bob through the key. If more information is embedded, the image is easily distorted. In contrast, if there is less embedded information, the image maintains good visual integrity, but does not meet our requirements for steganographic capacity. In this paper, we focus on tackling these challenges and limitations to improve steganographic capacity. An image steganography method based on Fully Convolutional Dense Network(FC-DenseNet) was proposed by us. The hidden network and the extracted network are trained at the same time. The dataset of the deep neural network is derived from various natural images of ImageNet. The experimental results show that the stego-image after steganography and the secret image extracted from stego-imge have a visually good effect, and the stego-image has high capacity and high peak signal to noise ratio. Image-to-image full size hiding is implemented.

## Ⅰ. INTRODUCTION

Information hiding is often associated with criminal activity, for example, some illegal information is hidden in the public image database to coordinate criminal activities [1]. Beyond the multitude of misuses, it can also be used in some practically good ways, such as image owner adds watermark to protect personal copyright in the image [2,3,4]. Information hiding [5,6] is mainly used for private communication between the two parties in online media. The goal of encryption [7] and steganography is to not discover the existence of secret information by third parties. Steganography is an important algorithm on the common channel in transmission, usually use text, images, audios or video as carrier secret information in the communication process, the carrier is a container that loads secret information and the carrier is also called stego-image after steganography.

Image steganography is the insertion of secret information into a carrier image. The message is embedded in the (LSB) [8,9] of the image in early image steganography, (LSB) digit of the pixel is easily replaced by secret information, which is easy to produce visual artifacts. The attacker finds that the secret information is easy in the image. In contrast, if the attacker knows the carrier, the contrast between the stego-image and the carrier image will be significantly different. An illegal attacker knows that information is embedded in the image. In [10], Xin Liao at al. proposes a medical JPEG image steganography scheme based on block-to-block coefficient dependence. The basic strategy is to keep the difference between DCT coefficients at the same position in adjacent DCT blocks as much as possible. The cost value is dynamically allocated based on the modification of the inter-block neighbors in the embedding process. In [11], Wei-Hung Lin at al. proposes a blind watermarking algorithm based on the significant difference between wavelet coefficient quantization and copyright protection. Each of the seven non-overlapping wavelet synergies of the carrier image is grouped into one block. It mainly quantifies the local maximum coefficient to determine the watermark. In [12], Yuan Chengsheng at al. propose a new coverless image steganography scheme based on scale-invariant feature transform and feature package. The basic scheme is that the secret information is converted into a bit stream by constructing a hash sequence known to the sender and the receiver. Interestingly, deep learning can hide information in the image, and our eyes do not visually see changes in the image. Converting and changing or modifying the frequency and spatial domains of carrier images in [8,9,10,11,12] to achieve steganography of text information. In recent years, as deep learning [13, 14] has become hotter and hotter, it has also contributed to the advancement of steganography. Text information is hidden into the image in [15,16,17], in adversarial network, the stego-images generated by the generator are such that the discriminator does not distinguish between authenticity and the carrier image in [16] and [17], at the same time, it is very difficult to distinguish between carrier image or stego-image in our vision. So we can't find secret information in the image.

Reversible information hiding is a popular trend today, and it is the ability to recover secret information to a certain extent. Weng S at al. proposes a reversible watermarking scheme based on local smoothness estimation and multi-step embedding strategy [18], all pixels in the picture are divided into 4 equal parts, corresponding to this watermark embedding is divided into four separate steps, and for each pixel to be embedded, the total variance of all its neighboring pixels is calculated to estimate the local smoothness. At low embedding rates, the pixels are only modified in the smooth region, however, adaptive embedding is applied when the low embedding rate increases. In [19], Qin C at al. propose a reversible new data hiding scheme based on the development direction (EMD). The main idea is to select a carrier image and generate two visually similar stego-image. The pixel value of the first image is modified by no more than one gray level to embed the secret information. Secondly, the second image is adaptively modified by referring to the first stego-image, and finally extracting the secret information from the two stego-images. Most of them are steganographically textual [8,9.10.11,15,16,17,20], and the number of bits in the text occupying each pixel is relatively small in the image, however, this method showed good resistance to existence discovery. A good information hiding challenge arises because the appearance and underlying statistical changes of the carrier image are easily caused by embedding the message. The amount of change depends on two factors: first, the amount of information that will be hidden. The most common is to hide relatively few bits of text information hidden, such as in [21]. Second, the amount of visible change depends on the carrier image itself. Hiding secret information in the high frequency region of the image is better than hiding the information in artificially detectable low frequency regions. The capacity for estimating information hiding can be found in [22]. In [23,24,25], the text is converted to binary data, and then the position of the LSB is automatically selected by the neural network for embedding. In contrast, in our work, the difference between us and [26] is that the preparation network is removed, and the hidden network uses FC-DenseNet [27]. The goal is the same as [26] to achieve the secret information fully embedded in the carrier image (the same secret image size as the carrier image). The secret image is scattered throughout all the bits and color channels in the pixels surrounding the carrier image. Embedding a complete $M \times N \times D$ secret image by using a carrier image of $M \times N \times D$ ($D = 3$, $D$ is the number of image channels), rather than simply modifying the LSB, and the carrier image has only a small distortion on each pixel.

Perhaps the best neural network is used at the same time as the work presented here [28]-[31]. In addition, in standard steganography studies, these methods encode small amounts of information, but are visually of good quality. Though similar conceptually to steganography [8]-[12],[18]-[20], four key differences set this work apart:

-- In [27], the image is segmented by semantics, and the segmentation effect is very good. Unlike [27], we applied FC-DenseNet for information hiding for the first time. The number of input channels of the first convolution block is 6, the number of output channels of the last convolution block is 3. In addition, we did not mark the different objects in the image, but deleted the label category and the 1×1 convolution.

-- We implicitly simulate the statistical mathematical distribution of natural images, rather than creating a mathematical model of the display, which is achieved by training a large number of cover and secret images through deep neural networks.

-- Hidden secret information does not need to be perfectly coded and can accept small errors. It can clearly balance the reconstruction quality of the carrier image and the secret image, as shown in Figure 1.

-- Unlike an encrypted noise image or an image that is obviously visually impacted after adding secret information. Instead of the images we transmit are meaningful images, and large orders of secret information are hidden, the ratio of the magnitude of the secret image to be hidden to the image of the carrier is 1:1.

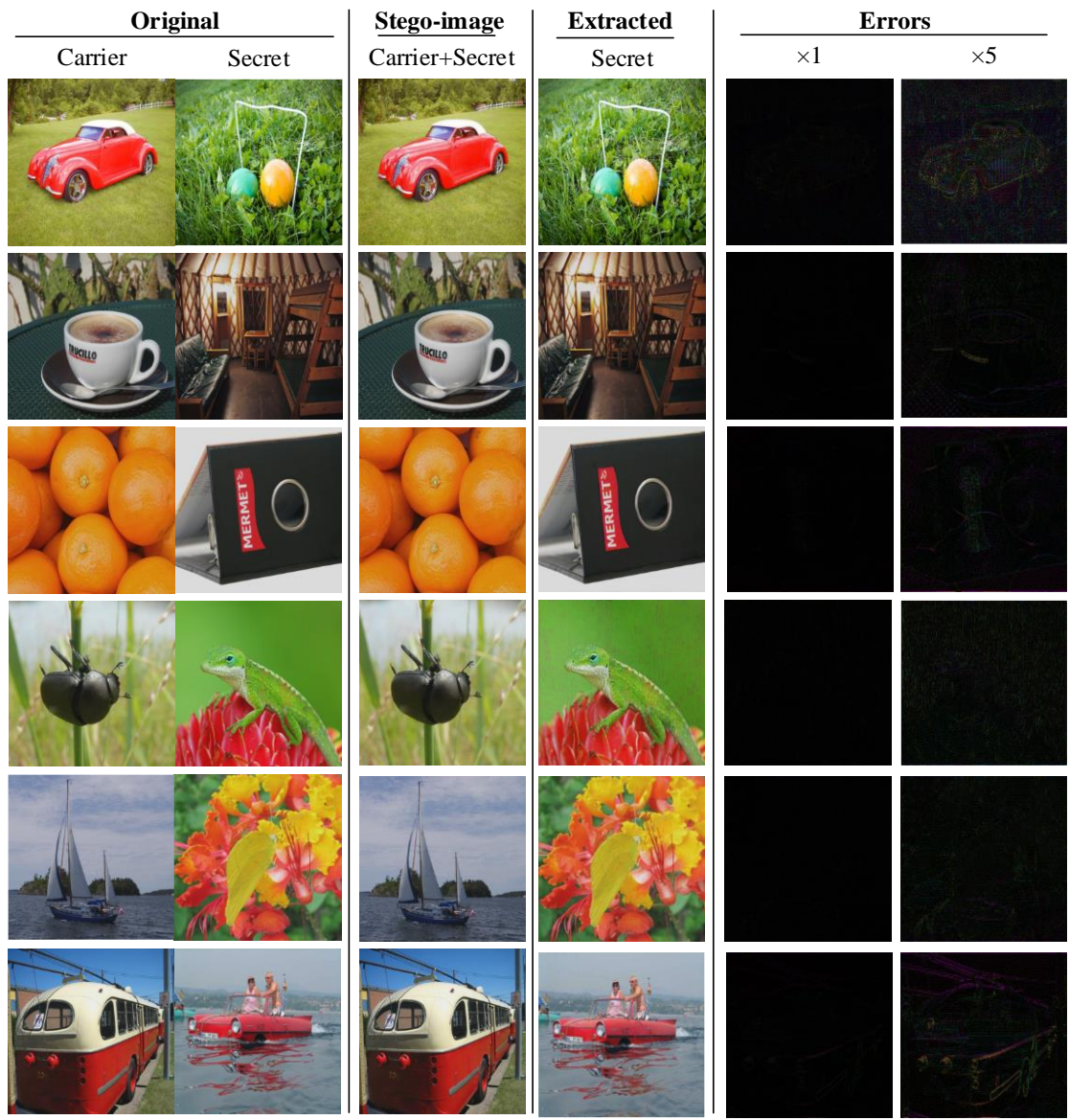

Fig. 1. Samples from the full-image hiding system. From the left: carrier image, secret image, the stego-image (the stego-image holds/hides the secrte image within it while looking like the carrier), and the recovered secret image – this is extracted from only the stego-image. The last two columns are the errors for the setgo-image vs. carrier (enhanced 1×, 5×).

## II RELATED WORKS

Deep networks have achieved unprecedented success in many image segmentation task such as [27],[32],[33]. In this article, we mainly apply FC-DensNet to our hidden network to hide information. FC-DenseNet is extended from DenseNet [34]. In order to utilize deep networks in image steganography, we will briefly introduce the application of Dense and FC-DenseNet to lay the foundation for the next section.

1 Dense Convolutional Network (DenseNet)

DenseNet is not only used for the classification of data but also for the super-resolution of images and image segmentation. In [35], Tong Tong al. proposed a method to apply DenseNet to image super-resolution. After the image passes DenseNet, it is followed by deconvolution and reconstruction, which has achieved good results. In view of the fact that the traditional convolutional network has an *L* layer, there are *L* number of connections. However, there is a connection between each layer of the DenseNet network and each subsequent layer, and the total number of connections is *L(L+1)/2*.

The advantage of DenseNet is to eliminate the vanishing-gradient problem, enhance feature propagation, encourage feature reuse, and significantly reduce the number of parameters. The structure of DenseNet is shown in Figure 2. Figure 2 briefly illustrates the layout of DenseNet. Consequently, the $l^{th}$ layers accept all layers in front of feature-maps, $X_0, X_1, \ldots, X_{l-1}$ as input:

$$X_l = Y_l([X_0, X_1, ..., X_{l-1}]) \quad (1)$$

where $[X_0, X_1, \ldots, X_{l-1}]$ means to the concatenation of the feature-maps produced in layers 0, 1, 2, …, *l-1*. Combine multiple tensors into one tensor for better application. *Y* is a composite function with three operations: namely batch normalization (BN)and rectified liner unit(ReLU), convolution(Conv).

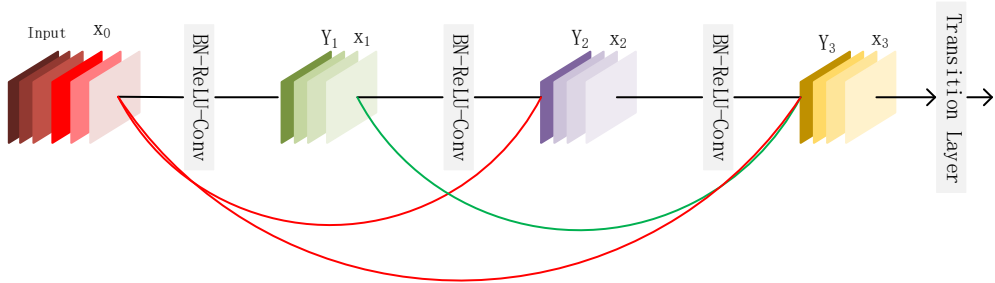

Fig. 2. A 4-layer dense block with hyperparameter k=4. The hyperparameter determines how much information each layer contributes globally.

2 Application and Development of FC-DenseNet

FC-DenseNet [36] as the name suggests is a dense convolutional neural network. It is derived from DenseNet, FC-DenseNet is mainly used for semantic segmentation, according to the label, it can segment the sky, tree, vehicle, pedestrian, road, etc. in the picture, and also applies to the segmentation in the video. The main process is to add a downsampling and upsampling function behind the Dense Block. Downsampling is a 2×2 convolution block used to extract the maximum value of the feature map to reduce the image and simplify the calculation. Upsampling is composed of a 3×3 transposed convolution with stride 2 to compensate for the pooling operation. Finally, the classification is done by 1×1 convolution and the number of categories. More straightforward, the number of output feature maps is the number of different categories, and the effect of the segmentation is attractive.

Based on FC-DenseNet, it is not necessary to split different objects in the image but to hide information in our work. We did not mark the different objects in the image, deleted the label category and the 1×1 convolution to achieve perfect hiding of the information.

In the next section, we will describe how to train both the encoder and the decoder to hide secret pictures and recover secret pictures, as well as the structure of the encoder and decoder. Quantitative assessment and visual assessment will be carried out in Section IV. Conclusions are present in Section V.

**III Proposed Image Steganography Scheme**

1 Network Architecture

As illustrated Fig. 3, our proposed architecture mainly contains concatenation and a encoder(Hiding Network) and a decoder(Reveal Network). Concatenation can be described as: let $c \in R^{W \times H \times D}$ and $s \in R^{W \times H \times D'}$ be two tensors of the same width and height and depth, $D$ and $D'$, where c and s are the carrier image and the secret image, respectively; then, let concatenation : $(c, s) \to \varphi \in R^{W \times H \times (D+D')}$ be the concatenation of the two tensors along the depth axis. The encoder is mainly used to hide secret information and the decoder is mainly used to extract secret information. Our goal is to train encoders and decoders at the same time to achieve perfect information hiding and extraction.

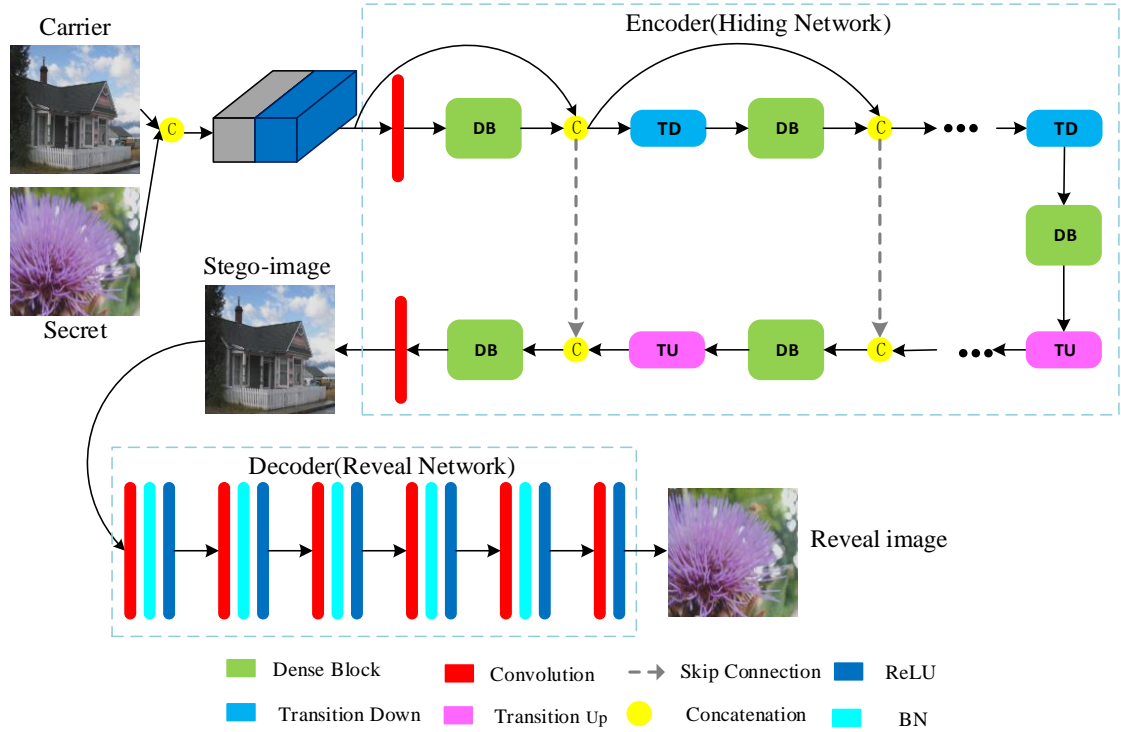

Fig. 3. Illustration of our model. The whole framework involves an encoder, a decoder, the input to the encoder is a 6-channel tensor and the output is a 3-channel tensor. The input to the decoder is the output of the encoder, and the output of the decoder is a 3-channel tensor.

2 Hide Encoder of secret image

The encoder was previously a concatenation operation. This function spliced the carrier image and the secret image and output a 6-channel image to prepare for the next full-size image hiding. The structure of the encoder is shown in Figure 3. The input to the encoder is a 6-channel tensor and the output is a 3-channel tensor. The main goal of the encoder is to encode a tensor with a channel number of 6 and the output stego-image should be as close as possible to the carrier image. Encoder components include two 3×3 convolution, 11 Dense Block(DB), 10 Concatenation, 5 Transition Down(TD) and 5 Transition Up(TU). The number of input channels of the first convolution block is 6, the number of output channels is 48, the number of input channels of the last convolution block is 192, and the number of output channels is 3. TD is composed of batch normalization, Relu, 1×1 convolution, 2×2 maximum pooling. TU is a 3 × 3 Transposed Convolution stride = 2. In Figure 4, each Dense Block (DB) consists of 4 layers, each layer including batch normalization, activation, 3 × 3 convolution, Droupt=0. Our goal is to minimize the loss between stego-image and carrier image:

$$\tau = \|c - c'\| \quad (2)$$

where c and c' represent the carrier image and stego-image.

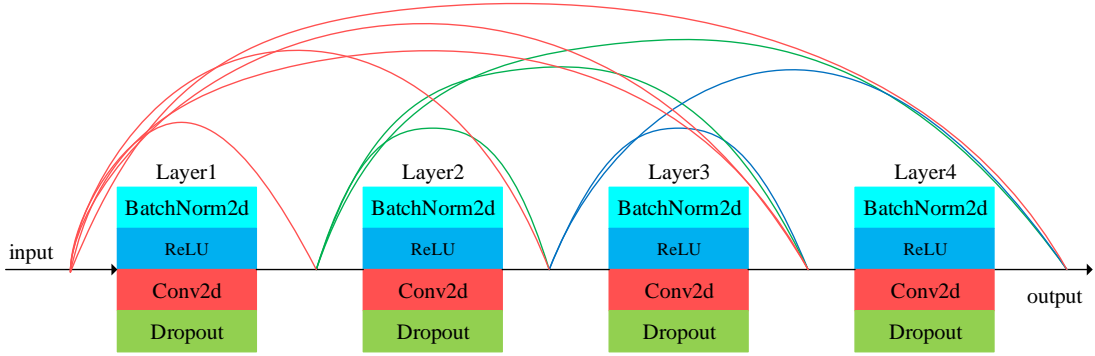

Fig. 4. Diagram of a dense block of 4 layers. Layers and layers are connected to each other

3 Reveal Decoder of stego-image

In Figure 3, the input to the decoder is a 3-channel tensor and the output is also a 3-channel tensor. The decoder components include 6 convolutions, 5 batch normalizations (BN), and 6 Relu. The main goal of the decoder is to decode the secret image (s') from the stego-image and the decoded s' should be as similar as the original secret image as possible. Our goal in the decoder is to minimize the loss of the decoded secret image and the original secret image:

$$\gamma = \|s - s'\| \quad (3)$$

where $s$ and $s'$ represent the original secret image and the reveal secret image, respectively.

4 Loss function

In our work, the target of the encoder is to hide the secret image into the carrier image to get the stego-image, the decoder extracts secret images from stego-image. The system is trained by reducing the following error: ($\beta$ is their weights):

$$\zeta = \tau + \beta\gamma \quad (4)$$

Notice that the error $\tau$ does not change the weights of the decoder. Because the decoder does not need to reconstruct the cover image; it only needs to recover secret information from stego-image. Both the encoder and the decoder are trained from the error signal $\beta\gamma$ since the hidden network and the extracted network are responsible for saving and forwarding information about hidden images. Better hiding and extracting information by propagating error signals.

IV Experiments

In this paper, 20,000 images are used for training, 3500 images are tested, and the dataset is from ImageNet. The initial learning rate of the network is set to 0.001, and the hyperparameter

β=0.75. The batch into which the image enters the model is set to 16, and the number of iterations of the training is set to 200. In the GPU is NVIDIA GeForce 1080, the experimental environment is Pytorch1.1.0, and the application is Python 3.6 for simulation experiments, the GPUs number is 2. For performance evaluation, we used two assessment methods: visual evaluation and quantitative evaluation. The image distortion is evaluated by the Structural Similarity (SSIM) and the Peak Signal-to-Noise Ratio (PSNR).

1 Visual Evaluation

In order to verify the performance of the model, we performed a human visual assessment of the carrier image, the secret image, and the stego-image. In Figure 5, after comparison, our eyes can't see the difference between the carrier image and the stego-image. The stego-image image is subtracted from the carrier image to obtain an error image. In the 1, 2, 3, 5 and 7 lines of Fig. 5, after the error image is magnified 10 times and 20 times, we can only see artifacts of the carrier image without seeing the secret image artifacts.

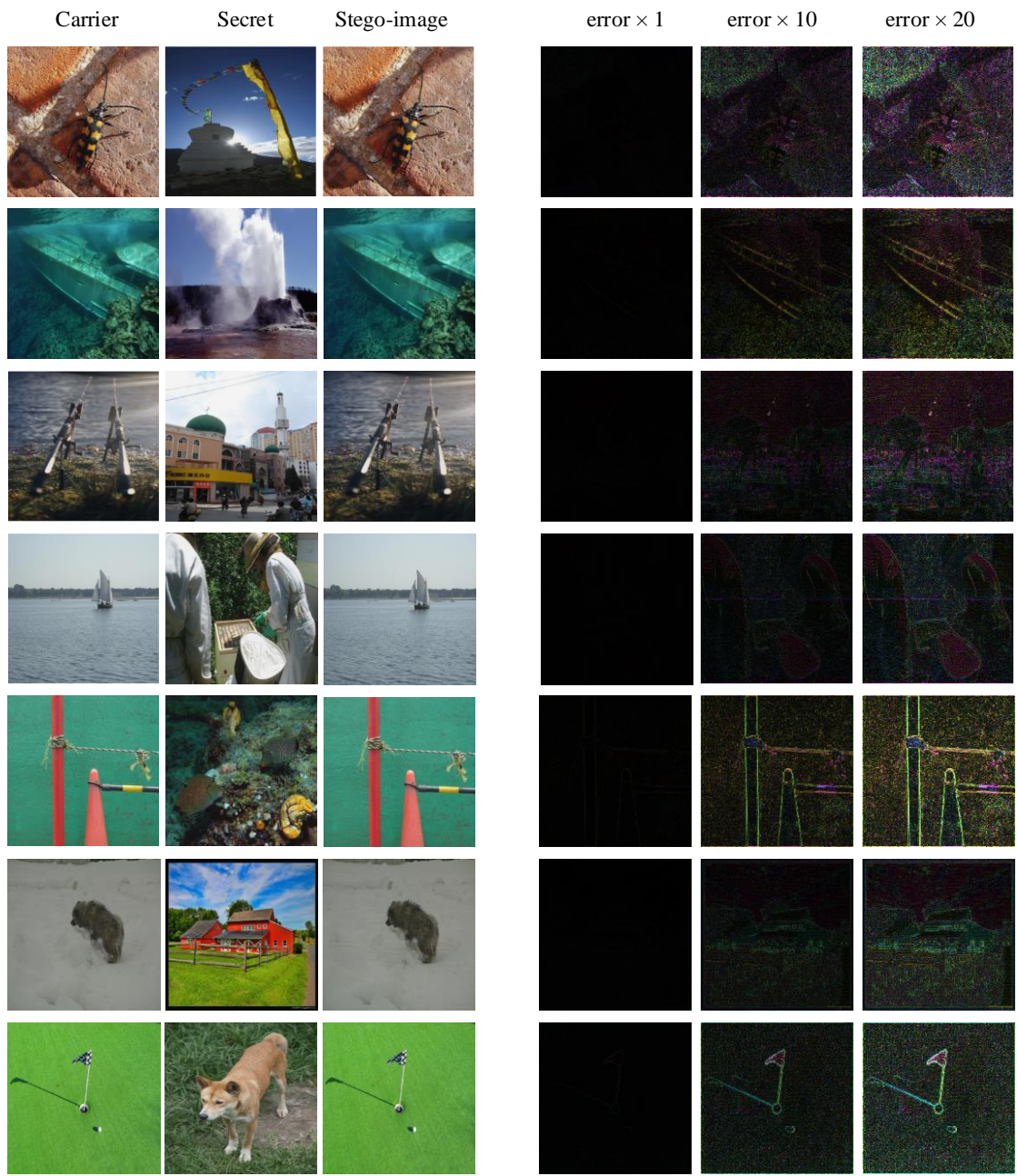

Fig. 5. From the left: carrier image, secret image, the stego-image. The last three columns are the errors for the setgo-image vs. carrier (enhanced 1×, 10×, 20×).

In order to further verify the generalization ability of the model, we selected several images from the CelebA dataset and the COCO dataset for testing. The results of the test are shown in Figure 6. The high frequency region of the carrier image has strong anti-interference ability, so the artifact of the carrier image appears in the error image. In the 4th line and the 6th line of Fig. 5, in the 7th line of Fig. 6, after the error image is magnified 10 times and 20 times, we only see the artifacts of the secret image blurred, because the area of the carrier image is relatively flat.

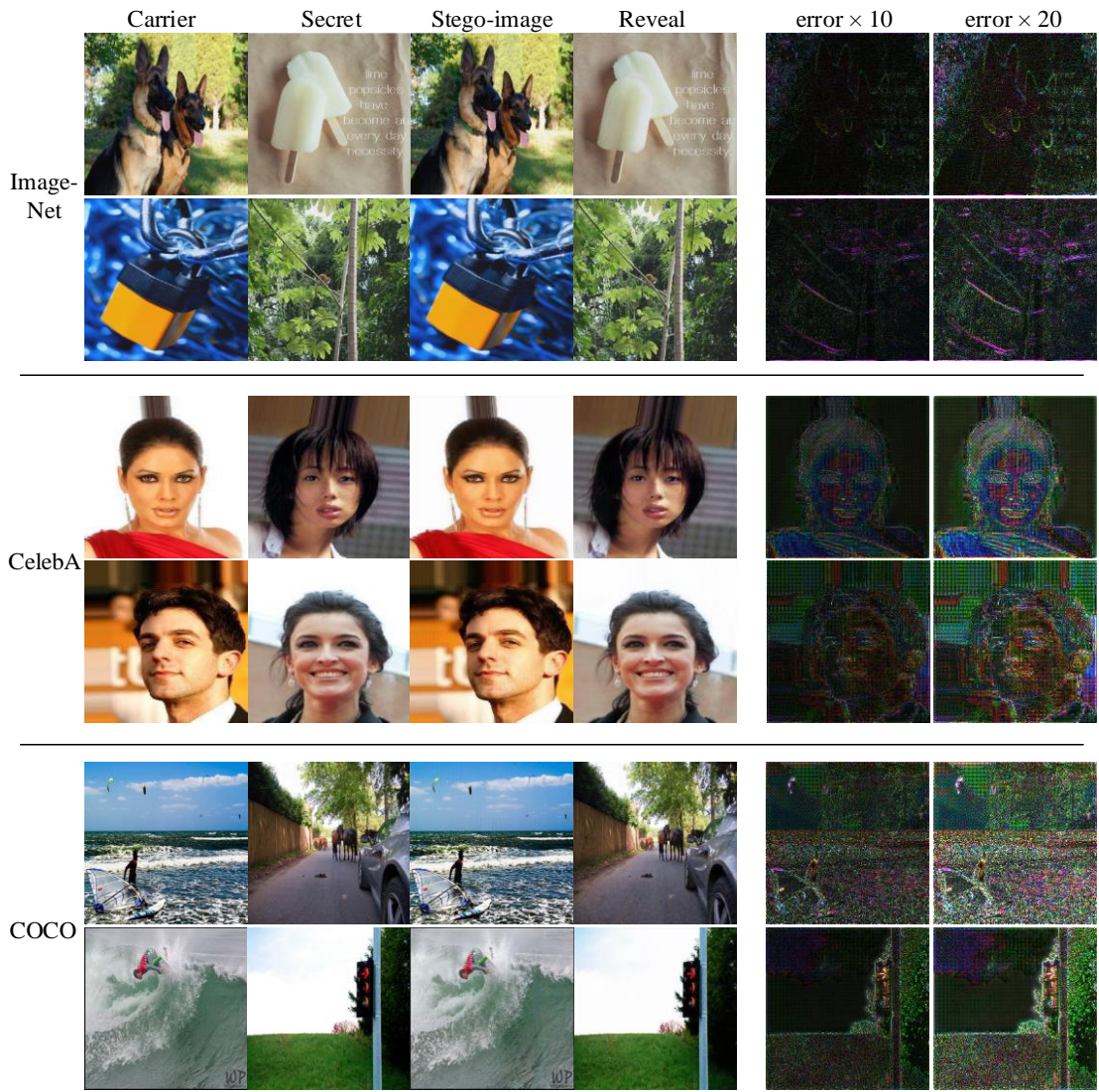

Fig. 6. From the left: carrier image, secret image, the stego-image (the stego-image holds/hides the secret image within it while looking like the carrier), and the recovered secret image – this is extracted from only the stego-image. 1 row and 2 rows, 3 rows and 4 rows, 5 rows and 6 rows respectively from ImageNet dataset, CelebA dataset, COCO dataset.

2 Quantitative Evaluation

In addition to visual assessment, we also need to measure the quality of stego-image. One widely-used metric for measuring image quality is the peak signal-to-noise ratio(PSNR), PSNR is mainly used to measure the distortion rate of an image and display it as a score. Given two images $X$ and $Y$ of size(W, H), the PSNR is defined as a function of the mean squared error(MSE):

$$MSE = \frac{1}{WH} \sum_{i=1}^{w} \sum_{j=1}^{H} (X_{i,j} - Y_{i,j})^2 \quad (5)$$

$$PSNR = 10 \log_{10} \frac{(2^n - 1)^2}{MSE} \quad (6)$$

where $(2^n - 1)^2$ is the square of the maximum value of the signal, and n is the number of bits of each sample value.

For a more complete evaluation of stego-image, we report the structural similarity index(SSIM) between the carrier image and the stego-image. Given two images $X$ and $Y$, the SSIM could be computed using the means, $\mu_X$ and $\mu_Y$, Variances, $\sigma_X^2$ and $\sigma_Y^2$, and covariance $\sigma_{XY}^2$ of the images as show below:

$$SSIM = \frac{(2\mu_X \mu_Y + k_1 R)(2\sigma_{XY} + k_2 R)}{(\mu_X^2 + \mu_Y^2 + k_1 R)(\sigma_X^2 + \sigma_Y^2 + k_2 R)} \quad (7)$$

here by default $k_1 = 0.01$, $k_2 = 0.03$, and the return value is between [-1, 1], where 1.0 means the two images are identical.

To further illustrate the effectiveness of steganography, in Table 1, we introduce the carrier image and the stego-image, the extracted secret image and the PSNR and SSIM of the original secret image. In the third row of Table 1, we can see that the average of PSNR and SSIM for the carrier image and the stego-images reached (40.451, 0.985), and the average of PSNR and SSIM of the reconstructed secret image and the original secret image reached (37.321, 0.981). In addition, in Figure 10, we plot the carrier image, stego-image, secret image, and the histogram of the reconstructed secret image, we can observe the carrier image and the stego-image, the secret original image and the reconstructed secret image, and the histogram difference between them is not very large. It is worth noting that because these evaluation indicators are relatively good, the solution we designed will not destroy the visual integrity.

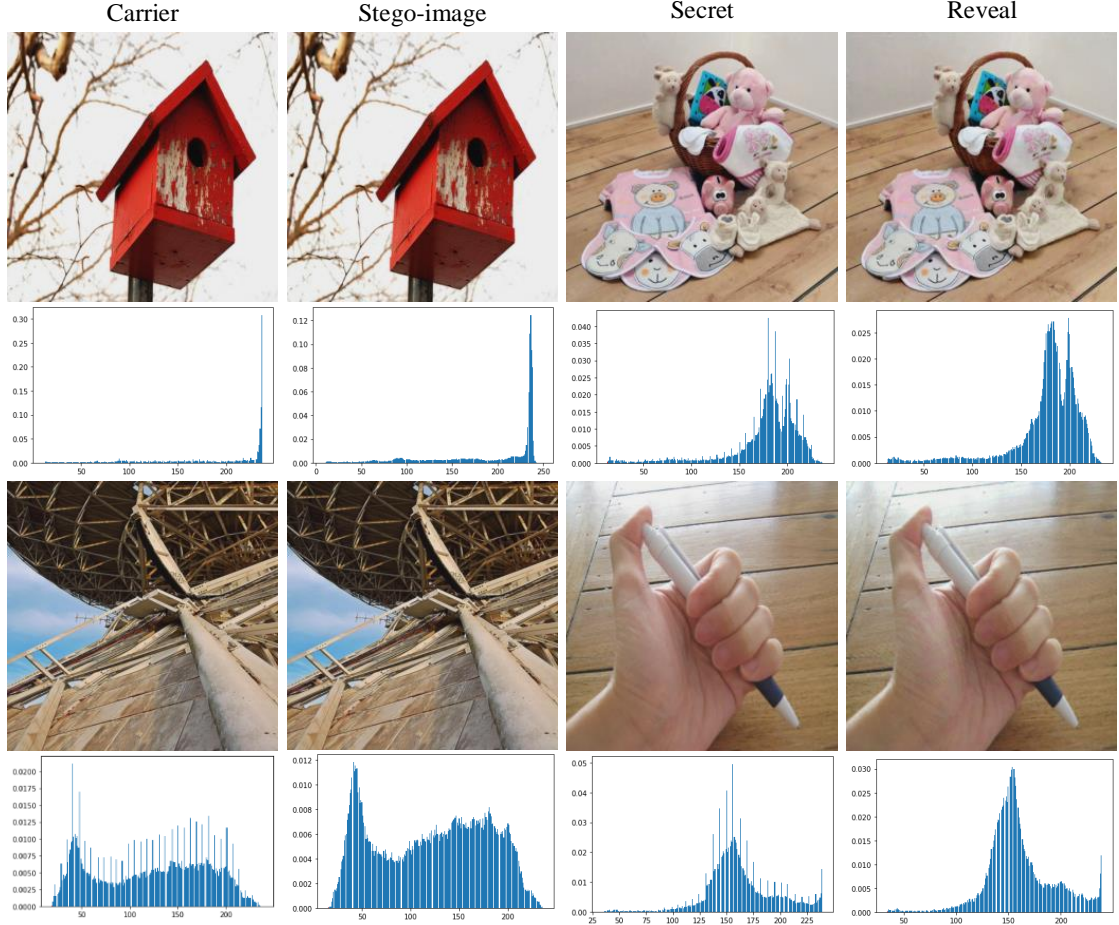

Fig. 7. From the left: carrier image, stego-image, secret image, and the recovered secret image. The 2nd and 4th line histograms correspond to lines 1 and 3, respectively.

TABLE 1

COMPARE CARRIER IMAGE AND STEGO-IMAGE, SECRET IMAGE AND RECONSTRUCTED SECRET IMAGE PSNR AND SSIM

|  | Carrier vs. stego-image (PSNR, SSIM) | Reconstructed secret vs. secret (PSNR, SSIM) |
|---|---|---|
| Fig.7 row1 | 40.196, 0.995 | 36.185, 0.984 |
| Fig.7 row3 | 39.274, 0.998 | 37.230, 0.974 |
| ImageNet (Average) | 40.451, 0.985 | 37.321, 0.981 |

3 Steganography Capacity Analysis

The traditional LSB steganography has relatively small steganographic capacity. Since our steganography scheme is relatively new, it is more intuitive to compare with other schemes, there are non-embedding hiding scheme including the carrier-selection-based and the carrier-synthesis-based methods. The results of the comparison are shown in Table 2, we can clearly see that the steganographic capacity of our proposed scheme is better than other schemes. Here, column 1, column 2, column 3, column 4 are steganography, steganography capacity per image, stego-image size, relative steganography capacity (steganography capacity per pixel ):

$$\text{Relative capacity} = \frac{\text{Absolute capacity}}{\text{The size of the image}} \quad (8)$$

TABLE2

Steganographic capacity comparison result

| Schemes | Absolute capacity (bytes/image) | Stego-image size | Relative capacity (bytes/pixel) |
|---|---|---|---|
| [37] | 1.125 | 512×512 | 4.29e-6 |
| [38] | 3.72 | ≥512×512 | 1.423-5 |
| [39] | 1533~4300 | 1024×1024 | 1.463-3~4.103-3 |
| Ours | 256×256 | 256×256 | 1 |

## V Conclusions

This paper is different from the traditional LSB steganography scheme, but through the end-to-end training model, the full-size image is hidden and the image has less distortion. Experimental results show that our method has advanced visual effects and high steganographic capacity. In the next step in this paper we will combine compression encryption and transmit only the parameters of the trained encoder to the receiver for secure and fast transmission.